\documentstyle[12pt]{article}
\newcommand{\sect}[1]{\setcounter{equation}{0}\section{#1}}

\begin{document} \def\bq{\begin{equation}}
\def\eq{\end{equation}}
\begin{flushright}
LPTENS/98/16
\end{flushright}

\begin{center}
{\large\bf
New phenomena in  the random field Ising model }
\end{center}

\begin{center}
{\bf E. Br\'ezin$^{a)}$ and C. De Dominicis$^{b)}$} \end{center} \vskip 2mm
\begin{center}{$^{a)}$ Laboratoire de Physique Th\'eorique, Ecole Normale
Sup\'erieure}\\ {24 rue Lhomond 75231, Paris Cedex 05,
 France{\footnote{ Unit\'e propre 701 du
Centre national de la recherche scientifique, associ\'ee \`a l'Ecole
Normale Sup\'erieure et \`a l'Universit\'e de Paris-Sud} } }\\
{$^{b)}$ Service de physique th\'eorique}\\ {Saclay, 91190 Gif-sur-Yvette,
France}\\
\end{center}
\vskip 3mm
\begin{abstract}
 We reconsider Ising spins in a Gaussian random field within the replica
formalism. The corresponding continuum model
involves several coupling constants beyond the single one which was
considered in the standard $\phi^4$ theory approach.
These terms involve more than one replica, and therefore in a mean field
theory they do not contribute to the zero-replica
limit. However the fluctuations involving those extra terms are singular on
the Curie line below eight dimensions, and by the time
one reaches the dimension six, it is necessary to keep them in the
renormalization group analysis. As a result it is found
that there is no stable fixed point of order $(6-d)$. Whether this means
that there is no expansion in powers of $(6-d)$,
or that the transition is driven to first order by these fluctuations, is
difficult to decide at this level, but it explains
the failure of the $(d, d-2)$ correspondence.
\end{abstract}
\newpage

In spite of more than twenty years of active and continuous research,
the simplest disordered system, the random field Ising model (RFIM),
is still far from a definite understanding.  The initial
(d, d-2) correspondence, between the RFIM in dimension d and the pure
system in dimension (d-2)(\cite{AIM,Young, Parisi-Sourlas}),
was suspected from the very beginning to break down in lower dimensions, if
one followed the arguments of Imry and Ma  (\cite{Imry}). The
correspondence was
definitely killed  when the mathematical approach of Imbrie, and
subsequently Bricmont and Kupiainen, finally established the existence of
an ordered state at
low temperature in three
dimensions(\cite{Imbrie,Bricmont}). But this still left room for various
possibilities. The elegant "derivation" of  the (d, d-2) relation by
 Parisi and Sourlas (\cite{Parisi-Sourlas}) made it clear that
non-perturbative issues could destroy the essential underlying supersymmmetry
 (since a Jacobian, the absolute value of a
determinant,  was replaced by the determinant itself which could change
sign). One could thus imagine that the (d, d-2)
correspondence for the critical exponents, although true to all orders in
an expansion
 in powers of
 $\epsilon = 6-d$, was modified by terms, such as $\exp- 1/\epsilon$, which
might very well turn out to be quite significant at
$\epsilon=3$. However progressive evidence was provided that the situation
could be more serious than that. Numerical
experiments in the supposedly paramagnetic phase, pointing at large time
scales before one reached the ordering temperature,
as well as the theoretical approach of M\'ezard and Young (\cite{Mezard}),
led to the idea that the phase diagram itself had to
be reconsidered. Coming from the high temperature regime it seems that one
meets a glassy phase, marked by replica symmetry
breaking (RSB), in a region in which there is no magnetic ordering, as in a
spin glass.
 We shall present in a subsequent
publication arguments which substantiate this analysis. However here we
argue that within the original models of Ising
spins, or of the
$\phi^4$ theory, in a Gaussian random field, another scenario might take
place. Indeed we show, within the replica approach,
that new interaction terms should be considered. These terms naively
disappear in the zero replica limit, but it is argued
that below dimension eight, these new interactions develop singularities
near the Curie line  in the $n=0$ limit, which change the analysis when one
reaches the dimension six.  The critical renormalization group flow, if
performed as usual on the Curie line,
 contains then five coupling constants, and it is easy to
show that there is no stable fixed point of order (6-d). Above the Curie
line these singularities
are cut-off by the finiteness of the correlation length, and there is a
non-uniformity in the approach of the Curie
line, versus the zero-replica limit. We are not able to say whether the
absence of a stable fixed-point means that there
is a non-perturbative fixed point elsewhere or whether the transition
becomes first order, but it shows that the dimensional reduction
may break down even at lowest order.
\sect{ The basic action} 
We consider successively, within the replica formalism,
 $\phi^4$ and Ising spins in an external Gaussian random field $h(x)$ with
\begin{equation}\label{1.1}
<h(x)> = 0 ; < h(x) h(y)> = \Delta \delta(x-y). \end{equation}
 Therefore starting with
	\begin{equation}\label{1.2}
 S(\phi) =\int dx [{1\over{2}} (\nabla\phi)^2 +{1\over{2}} r_0 \phi^2 +
{g\over{8}}
\phi^4 -h(x) \phi(x)]  \end{equation}
(the  Boltzmann weight is $exp-S$) and replicating $n$ times  we average
over the random field and obtain
\begin{equation}\label{1.3}
 S(\phi_\alpha) =\int dx [\sum_{\alpha=1}^{n} ({1\over{2}}
(\nabla\phi_{\alpha})^2 +{1\over{2}} r_0 \phi_{\alpha}^2 +
{g\over{8}}
\phi_{\alpha}^4) -
{\Delta\over{2}}\sum_{\alpha,\beta}\phi_{\alpha}(x)\phi_{\beta}(x) ]
\end{equation}
Let us introduce the notation
	\begin{equation}\label{1.4}
 \sigma_k = \sum_{\alpha=1}^{n} (\phi_{\alpha}(x))^k ; \end{equation}
the symmetry of this action under permutation of the replicas allows for five,
 a priori relevant, coupling constants  corresponding to the quartic
interactions $\sigma_4, \sigma_1\sigma_3, (\sigma_2)^2,
(\sigma_1)^2\sigma_2, (\sigma_1)^4$ .

Indeed if we were considering the renormalization programme
for the various correlation functions of the theory at finite n ,
 we would be forced  to introduce these coupling constants,
even if they are not all present initially. In order to substantiate this
claim, let us start with Ising spins and derive the
corresponding $\phi^4$ theory in the usual way.  We begin with

\begin{equation}\label{1.5}
 \beta H = -{1\over{2}} \sum_{i,j} J_{ij} S_i S_j -\sum_i h_iS_i ,
\end{equation}
and use the Hubbard-Stratonovich identity
\begin{equation}\label{1.6}
\exp[{1\over{2}} \sum {J_{ij} S_i S_j} +\sum {h_iS_i}] = C \int ({\prod_i}
{ d{\phi_i}}) \exp[-{1\over{2}}
\sum\phi_iJ^{-1}_{ij}
\phi_j +\sum (\phi_i +h_i)S_i] \end{equation} We then introduce n replicas
and sum over the spins. This gives the partition
function
\bq\label{1.7}
Z^n= C \int \prod_{i=1}^{N} \prod_{\alpha=1}^{n}d{\phi_{i,\alpha}}
\exp[-{1\over{2}}
\sum_{ij,\alpha}\phi_{i,\alpha}J^{-1}_{ij}\phi_{j,\alpha}
+\sum_{i,\alpha}\ln{\cosh(\phi_{i,\alpha}
+h_i)}]
\eq

We then expand in powers of $\phi$ , up to degree four:

\begin{equation}\label{1.8}
\ln {\cosh(\phi +h)\over\cosh {h}} = \ln ( 1 + \tau\phi+ {1\over{2}}\phi^2
+ {\tau\over{6}} \phi^3 +
{1\over{24}}\phi^4 + O(\phi^5))\end{equation}

in which $\tau = \tanh h$, and average over the random field $h$:
\begin{eqnarray}\label{1.9}
<\exp\sum_{\alpha} \ln{\cosh(\phi_{\alpha}+h)\over{\cosh h}}>= \exp(
{1\over{2}}(1-\tau_2)\sigma_2 +
{\tau_2\over{2}} \sigma_1^2 \nonumber\\ -{1\over{12}} (1 -4\tau_2 +
3\tau_4) \sigma_4 + {1\over {24}}(\tau_4
-3\tau_2^2)\sigma_1^4 +  {1\over {8}}(\tau_4 -\tau_2^2)\sigma_2^2
\nonumber\\+ {1\over{3}} (\tau_4 -\tau_2)\sigma_1\sigma_3
-  {1\over{4}} (\tau_4 -\tau_2^2)\sigma_1^2\sigma_2 + O(\phi^6))
\end{eqnarray}

with
\begin{equation}\label{1.10}
\tau_p = {1\over {\sqrt{2\pi\Delta}}}
\int_{-\infty}^{+\infty}dhe^{-{h^2\over{2\Delta}}} (\cosh{h})^n(\tanh{h})^p
\end{equation}

At the level of the quadratic terms  $(1-\tau_2)\sigma_2$
shifts the constant interaction term $\tilde{J}^{-1}(q=0)$
                and gives an effective 'mass'. The other term $\tau_2 \sum
\phi_{\alpha}\phi_{\beta}$
is due to the average over the random field. For a small random field, i.e.
small $\Delta$ , in the $n=0$ limit
\begin{equation}\label{1.11}
\tau_2 = \Delta - 2\Delta^2 + O(\Delta^3)   ;   \tau_4= 3\Delta^2 -
20\Delta^3 + O(\Delta^4)
\end{equation}

We are then led to a   $\phi^4$- theory with five coupling constants:
\begin{eqnarray}\label{1.12}
\beta H = \int d^{d}x ( {1\over{2}}\sum_{\alpha} [(\nabla\phi_{\alpha})^2 +
t \phi_{\alpha}^2 ] - {\Delta\over{2}}
\sum_{\alpha,\beta} \phi_{\alpha} \phi_{\beta}\nonumber\\
+ {u_1\over{4!}}\sigma_4 + {u_2\over{3!}}\sigma_1\sigma_3 +
{u_3\over{8}}\sigma_2^2 + {u_4\over{4}}\sigma_1^2\sigma_2
+{u_5\over{4!}}\sigma_1^4)
\end{eqnarray}

A priori the four additional terms beyond the single interaction $\sigma_4$ ,
 which is the only one usually present, involve more than one single
replica. Therefore if one writes the quenched average of
the replicated partition function in terms of $W =\rm{ Log} Z$ ,
\begin{equation}\label{1.13}
<Z^n> = \exp[n<W> + {n^2\over{2}}(<W^2> - <W>^2) + O(n^3)],
\end{equation}
it is easily seen that the contribution of the four additional coupling
constants are of relative order $n$ (or higher) and thus
could be discarded.However this argument is true provided the coupling
constants $u_i$  for
$i>1$, are not singular in the $n= 0$ limit.
\sect{Singularities below dimension eight}

At the level of mean field theory the coupling constants that we have
obtained in (\ref{1.9}) are all finite in
the $n=0$ limit. However the situation changes with the fluctuations:
consider the one-loop fluctuations in the theory at the
critical point. These fluctuations are given by the determinant generated
by the Gaussian fluctuations around the mean field,
namely
\bq\label{2.1} \exp-{1\over{2}}\rm{Tr}\ln (1 + G_0S^{(2)})\eq
in which
\bq\label{2.2} G_0^{\alpha\beta}(q) = {\delta_{\alpha\beta}\over({q^2}+t)}
+ {\Delta\over{({q^2}+t)({q^2}+t-n\Delta)}} \eq
and $S^{(2)}$ is the second derivative of the action (\ref{1.12}) with
respect to $\phi_{\alpha}$ and $\phi_{\beta}$
\begin{eqnarray}\label{2.3} S^{(2)}_{\alpha\beta} = \delta_{\alpha\beta}
[{1\over{2}}u_1\phi_{\alpha}^2 + u_2 \sigma_1
\phi_{\alpha} + {1\over{2}}u_{3}\sigma_2 + {1\over{2}}u_{4}{\sigma_1}^2]
+{1\over{2}}u_2(\phi_{\alpha}^2 +\phi_{\beta}^2)
\nonumber\\ +u_3 \phi_{\alpha}\phi_{\beta} + {1\over{2}}u_4\sigma_2 +
u_4\sigma_1(\phi_{\alpha}+\phi_{\beta})
+{1\over{2}}u_5{\sigma_1}^2
 \end{eqnarray}
in which we have used the notation (\ref{1.4}).
The fluctuations terms  which are quartic in $\phi$ are obtained by
expanding (\ref{2.1}) to
second order in $S^{(2)}$. We first consider the most singular part in
which for the propagator $G_0$ we retain twice the
$\Delta$ part of (\ref{2.2}). It involves the sum
\begin{eqnarray}\label{2.3}
\Delta^2(\sum_{\alpha\beta}S^{(2)}_{\alpha\beta})^2 = \Delta^2[
{1\over{2}}(u_1+ 4nu_2+nu_3+n^2u_4)\sigma_2
\nonumber\\ +(u_2+u_3+{5\over{2}}nu_4+n^2u_5)\sigma_1^2]^2\end{eqnarray}
multiplied by the one-loop integral
\bq I_n(p,t) =\int d^{d}q
{1\over{({q^2}+t)({q^2}+t-n\Delta)((p-q)^{2}+t)((p-q)^2+t-n\Delta)}}.\eq

The contribution of these fluctuations to the effective coupling
constants are thus related to $I_n(0,t)$. Note that one cannot let $t$ go
to zero first,
since this would yield  a pole of the integrand on the line $q^2=n\Delta$.
Therefore in order to determine the critical behaviour of this
effective interaction we have to examine $I_n(0,t)$ for $t$ of order
$n\Delta$, and then let $n$ go to zero. On the line
$t=n\Delta$, which is close to the usual critical
$t=0$ line, the integral $I_n(0,t)$ becomes
singular, in dimensions lower than eight. A simple power counting gives
\bq I_n(0,n\Delta) = \int d^{d}q {1\over{{q^4}}({q^2}+n\Delta)^{2}}\sim
(n\Delta)^{-{8-d\over{2}}}.\eq
Therefore these fluctuations yield singular contribution to the
coupling constants contained in  (\ref{2.3})  in dimensions lower than
eight in the zero-replica limit.  Thus the coupling constant
$u_3$ receives a contribution proportional to ${1\over{n}}u_1^2$ in
dimension six, likewise for $u_2$. On the other
 hand $u_4$ behaves like $u_1u_3/n$, i.e. like $1/n^2$ and $u_5$ like
$u_3^2/n$, i.e. $1/n^3$.
Consequently it is not legitimate to discard the coupling constants which
couple several replicas
in dimensions smaller than eight.

It is interesting to analyze how a dynamical approach, based for instance
on a Langevin formalism, would recover
those singularities, despite the absence of replica. In the
Martin-Rose-Siggia formulation  of the dynamics (\cite{Martin}), there are
two fields,
namely $\phi(x,t)$ the order parameter, and the conjugate field
$\hat\phi(x,t)$. The equilibrium $\phi^4$ theory leads to
a single dynamical interaction term of the form $\hat\phi \phi^3$. However
fluctuations build up additional couplings such as
$\phi^2\hat\phi^2$ ,etc., with singular coefficients in dimensions lower
than eight. It is then necessary to introduce a waiting time and invariance
under
time-translation invariance is broken . The  $n=0$
singularity in the critical domain is then replaced by a singularity when
the waiting time becomes large. A full comparison between the dynamical
approach and the
equilibrium replica formalism goes beyond the scope of this letter.

 \sect{The renormalization group flow near six dimensions}

Since the multi-replica coupling constants are singular below dimension eight,
one cannot discard the new interaction terms in the zero-replica limit,
since they blow up as powers of $1/n$ in
dimension six. Therefore we  keep now the full action (\ref{1.12})
with the five coupling constants, and repeat the usual  renormalization
group approach near six dimensions. We know that if we had kept only the
coupling
constant $u_1$ we would have found a stable fixed point of order $\epsilon=
6-d$, which would yield
a critical behaviour identical to that of the pure system near dimension
four(\cite{Young, Parisi-Sourlas}).
At one-loop we obtain the renormalization group equations by returning to
(\ref{2.1}), expanding again to second order in
$S^{(2)}$, but now keeping only one $\Delta$ in the propagators. The result
is proportional to
\begin{eqnarray}\label{3.1}
\sum_{\alpha}(\sum_{\beta}S^{(2)}_{\alpha\beta})^2 = &&{1\over{4}}
(u_1+nu_2)^2\sigma_4 + (u_1+nu_2)(u_2+u_3+nu_4)\sigma_1\sigma_3 +
 \nonumber\\+&&{1\over{4}}[n(u_2+u_3+nu_4)^2 +
2(u_2+u_3+nu_4)(u_1+nu_2)]{\sigma_2}^2\nonumber\\
+&&[2(u_2+u_3+nu_4)^2 +
 {1\over{2}}(3u_4+nu_5)(u_1+2nu_2+ nu_3+
n^2u_4)]{\sigma_1}^2\sigma_2\nonumber\\
+&&[(u_2+u_3+nu_4)(3u_4+nu_5) + {n\over{4}}(u_4+nu_5)^2]\sigma_1^4\nonumber\\
 \end{eqnarray}
multiplied by an integral proportional to $1/\epsilon$.
Redefining the coupling constants to contain geometric factor ( surface of
the unit sphere divided
by $(2\pi)^d$) we obtain the five one-loop beta functions :
\begin{eqnarray}\label{3.2} \beta_1 &=& -\epsilon u_1 + 3\Delta
(u_1+nu_2)^2 \nonumber\\
 \beta_2 &=& -\epsilon u_2 + 3\Delta (u_1+nu_2)(u_2+u_3+nu_4) \nonumber\\
 \beta_3 &=&-\epsilon u_3 + \Delta[n(u_2+u_3+nu_4)^2 +
2(u_2+u_3+nu_4)(u_1+nu_2)]\nonumber\\
 \beta_4 &=& -\epsilon u_4 + \Delta[4(u_2+u_3+nu_4)^2\nonumber\\ &&+
n(u_2+u_3+nu_4)(3u_4+nu_5)+ (3u_4+nu_5)(u_1+nu_2)]\nonumber\\
\beta_5 &=& -\epsilon u_5 + 12\Delta[(u_2+u_3+nu_4)(3u_4+nu_5) +
{n\over{4}}(3u_4+nu_5)^2]\end{eqnarray}

These equations make it clear that the renormalization mixes the five
operators that we
are considering. If, notwithstanding the potential singularities of
$u_2,\cdots,u_5$, one lets $n$ go to zero first in the equations
\ref{3.2}, one recovers
the usual dimensional reduction fixed point $ u_1=
\frac{\epsilon}{3\Delta}+ O(\epsilon^2),
u_2=\cdots=u_5=0$. It is immediate to check that this fixed point is
unstable and that the $u_{i}'s$ for $i>2$ grow. Furthermore
we have seen in the previous section that the effective interactions with the
symmetry of those extra coupling constants, are indeed singular when $n$
goes to zero.
Therefore instead of letting $n$ go to zero first, we give to these
coupling constants the dependence in $n$ inherited from
these singularities, i.e.  we combine the $n$-dependence from these
singularities with the sum over replica indices.
Examining those beta functions \ref{3.2}, it is immediate to verify that if we
redefine the coupling constants
\bq\label{3.7} g_1= \Delta u_1, g_2= n\Delta u_2, g_3 = n\Delta u_3,
g_4=n^2\Delta u_4,g_5=n^3\Delta u_5 \eq
the number of replicas $n$ drops completely from the equations. Note that
the dependence in $n$ of these coupling constants is exactly
consistent with the analysis of the effect in dimension six of the
multi-replica singularities analyzed in the previous section.
 This leads to the important conclusion that the existence and
stability of the fixed points is now independent of the number of replicas.
In terms of the couplings $g_i$ we now have

\begin {eqnarray}\label{3.8} \beta_1 &=& -\epsilon g_1 + 3 (g_1+g_2)^2
\nonumber\\
 \beta_2 &=& -\epsilon g_2 + 3 (g_1+g_2)(g_2+g_3+g_4)\nonumber\\
 \beta_3 &=& -\epsilon g_3 + (g_2+g_3+g_4)^2 +
2(g_2+g_3+g_4)(g_1+g_2)\nonumber\\
 \beta_4 &=& -\epsilon g_4 + 4(g_2+g_3+g_4)^2 +
(g_2+g_3+g_4)(3g_4+g_5)+ (3g_4+g_5)(g_1+g_2)\nonumber\\
\beta_5 &=& -\epsilon g_5 + 12(g_2+g_3+g_4)(3g_4+g_5) + 3(3g_4+g_5)^2\end
{eqnarray}
(we have kept the same name to the beta functions, in spite of an obvious
refinition by a multiplicative factor
corresponding to the redefinition (\ref{3.7})).
We now look for fixed points $g_i*$ which are zeroes of those equations and
satisfy the stability condition
given by the non-negativity of the eigenvalues (or rather the real part of
the eigenvalues if they were complex)
of the matrix
\bq\label{3.12}  \Omega_{ij} = {\partial{\beta_i}\over{\partial{g_j}}}
\vert_{g*}\eq
at the fixed point. Note that if $g_1$ was the only coupling, the fixed
point $g_1* = {\epsilon\over{3}}$ would
be stable, but it is no longer true when the others are present since
${\partial{\beta_3}\over{\partial{g_3}}}= -{\epsilon\over{3}}$ being
negative, the matrix $\Omega$
cannot be non-negative.
The discussion of the fixed point solutions is easier in terms of the variables
\bq g_1+g_2 =x, g_2+g_3+g_4=y, 3g_4+g_5 =z .\eq
and the first equation (\ref{3.8}) leaves us with two possibilities
a) $x=0$ and then ${\partial{\beta_1}\over{\partial{g_1}}}$ is negative,
b) $x\neq 0$ , then $x+y = {\epsilon\over{3}}$ and
${\partial{\beta_3}\over{\partial{g_3}}}= -\epsilon +2(x+y)$
is negative. We then conclude again that there is no stable fixed point.
Therefore, whether we take into account
the singular dependence in $n$ of the effective interactions or not, one
does not find a stable fixed point of
order $(6-d)$.
This still leaves us with two possibilities (i) a stable fixed point which
is not of order $\epsilon$ (ii) a runaway flow
indicative of a first order transition. However in both cases there is no
epsilon-expansion of the critical properties of the
random field Ising model, and the dimensional reduction breaks down at
first order in $\epsilon$.
 \sect{Branched polymers}
One may wonder at this stage whether the previous analysis would not
endanger as well the beautiful
results of Parisi-Sourlas on dimensional reduction for branched
polymers(\cite{PS}), for which there are only reasons
to believe that it works. We shall not repeat here their original
derivation but simply take the replicated field
theory version of the problem, which is now a $\phi^3$ theory. The action
is similar to (\ref {1.12})
\bq\label{4.1}
\beta H = \int d^{d}x ( {1\over{2}}\sum_{\alpha} [(\nabla\phi_{\alpha})^2 +
t \phi_{\alpha}^2 ] - {\Delta\over{2}}
\sum_{\alpha,\beta} \phi_{\alpha} \phi_{\beta}
+ {u_1\over{3!}}\sigma_3 + {u_2\over{2}}\sigma_1\sigma_2 +
{u_3\over{3!}}\sigma_1^3).
\eq
Again the theory has been based on the single coupling constant $u_1$ since
the other two drop from the problem
in the zero-replica limit. The upper critical dimension is then eight, and
the authors of (\cite{PS}) proved that the
epsilon-expansion about eight dimensions (based on a purely imaginary
coupling constant) is identical to the $(6-d)$ expansion
for the $\phi^3$ theory.

Similarly here the multi-replica coupling constants develop singularities
in the zero-replica limit below dimension twelve
which make it necessary to reconsider the analysis of the
$(8-d)$-expansion. Following the same lines we have obtained
the three beta functions. Again $n$ drops from the determination fo the
fixed points if we redefine
\bq g_1= \Delta u_1,g_2 = n\Delta u_2, g_3= n^2\Delta u_3 \eq
and we find
\begin{eqnarray} \beta_1 &=&- \frac{\epsilon}{2} g_1 -2g_1 (g_1+g_2)^2
-\frac{1}{4}g_1(g_1+3g_2+g_3)^2\nonumber\\
\beta_2 &=& -\frac{\epsilon}{2} g_2 -\frac{1}{4}g_2(g_1+3g_2+g_3)^2 \nonumber\\
&&-2g_1(g_1+g_2)(2g_2+g_3)
 -2g_2(g_1+g_2)(g_1+3g_2+g_3)\nonumber\\
\beta_3 &=& -\frac{\epsilon}{2} g_3 + g_3 (g_1+g_2)^2 -3(2g_2+g_3)^2
(g_1+g_2)-6g_2(g_1+3g_2+g_3)(2g_2+g_3) \nonumber\\
&&-\frac{13}{4} g_3(g_1+3g_2+g_3)^2 -6g_2(g_1+g_2)(2g_2+g_3)
\end{eqnarray}
The Parisi-Sourlas fixed point is $g_1*^2 = -\frac{2\epsilon}{9},
g_2*=g_3*=0 $, and it is immediate to verify
that in this
case it is (marginally) stable. Therefore the mechanism which invalidates
the epsilon-expansion for the RFIM is harmless
for the problem of branched polymers.

\begin{center}
{\bf Acknowledgement}
\end{center}
It is a pleasure to acknowledge fruitful discussions with Marc M\'ezard and
Nicolas Sourlas, and we thank
 Giorgio Parisi for a useful correspondence,
during the course of this work.


\end{document}